\documentclass[preprint,preprintnumbers,amsmath,amssymb,nofootinbib]{revtex4}

\usepackage{etex}
\usepackage{amssymb,amsthm,amscd,amsbsy,array}
\usepackage{bm}
\usepackage{soul} 
\usepackage{graphics,graphicx,xcolor}
\usepackage{soul} 
\usepackage{algorithm}
\usepackage{algorithmic}
\usepackage{longtable}
\usepackage{graphicx}
\usepackage{epstopdf}
\usepackage{subfigure}
\usepackage{float}
\usepackage[utf8]{inputenc}
\usepackage[T1]{fontenc}
\usepackage[english]{babel}
\usepackage{graphicx}
\usepackage{float}
\usepackage{tikz}
\usepackage{xcolor}
\usepackage[utf8]{inputenc}
\usepackage[T1]{fontenc}
\usepackage[english]{babel}
\usepackage{graphicx}
\usepackage{float}
\usepackage{tikz}
\usepackage{xcolor}
\usepackage{makecell}
\usepackage{slashed}

\usepackage[colorlinks=true, pdfstartview=FitV, linkcolor=blue, citecolor=blue, urlcolor=blue]{hyperref} 

\begin{document}
		
	\title{Hidden symmetries of a self-dual monopole
	\footnote{Originally appeared
in: {\sl  Symmetries in Science III}, Proc. Schloss Hofen
 Meeting, Gruber and Iachello (eds.), pp. 525-529. Plenum (1989). Available in inspire (code {Feher:1988ym}).}
 }
	
	\author{
		L. Feher$^{1,2}$~\footnote{Present address lfeher@physx.u-szeged.hu},\,
		P. Horv\'athy$^{2,3}$~\footnote{Present address: horvathy@univ-tours.fr}\,
		and
		L. O'Raifeartaigh$^{2}$ \footnote{deceased}
	}
	
	\affiliation{
		${}^1$ Bolyai Institute, Jate H-6720 - Szeged (Hungary)
		\\
		${}^2$ Dublin Institute for Advanced Studies, - Dublin (Ireland)
		\\
		${}^3$ Laboratoire de Math\'ematiques et Applications de Metz (B\^at. A, Ile du Saulcy), Universit\'e Paul Verlaine - Metz (France)
	}
	\date{\today}		
	
\begin{abstract}
 The symmetries of a spinning particle in the field of a self-dual monopole are studied from the viewpoint of supersymmetric quantum mechanics.
\end{abstract}
	
	\maketitle
	
	
A self-dual (SD) $SU(2)$ monopole is a static solution of the first order Bogomolny \cite{ref-1} equations
\begin{eqnarray}
D_{k} \Phi=B_{k}\left(=\frac{1}{2} \epsilon_{k i j} F_{i j}\right).
\end{eqnarray}
The monopole field can be identified with a pure Yang-Mills configuration $A_{0}=0$, $A_{k}, A_{4}=\Phi$ in (1+4)-dimensional flat space which does not depend on the coordinates $x^{0}$ and $x^{4}$. The Dirac equation,
\begin{eqnarray}
	i \frac{\partial \Phi}{\partial x^0}=i \gamma^0 \slashed{D}_4 \Psi,
\end{eqnarray}
plays a decisive role \cite{ref-2,ref-3,ref-4} in describing the fluctuations around a SD monopole. Here we are interested in the symmetries of the 4-dimensional, Euclidean Dirac operator
\begin{eqnarray}
\slashed{D}_{4}=\gamma^{k} D_{k}+\gamma^{4} D_{4}=\frac{\sqrt{2}}{i}\left(\begin{array}{cc}
	0 & Q \\
	Q^{+} & 0
\end{array}\right),
\end{eqnarray}
where
\begin{eqnarray}
&&Q=\frac{1}{\sqrt{2}}\left(\Phi \cdot \mathbf{1}_{2}-i \vec{\pi} \cdot \vec{\sigma}\right),
\nonumber \\
&&\vec{\pi}=-i \vec{D}=-i \vec{\nabla}-\vec{A}^{a} I_{a}, \quad D_{4}=\frac{\partial}{\partial x^{4}}-i \Phi^{a} I_{a}.
\end{eqnarray}
We use the following $\gamma$-matrices:
\begin{eqnarray}
\gamma^{k}=\left(\begin{array}{cc}
	0 & i \sigma^{k} \\
	-i \sigma^{k} & 0
\end{array}\right), \quad \gamma^{4}=\left(\begin{array}{cc}
	0 & \mathbf{1}_{2} \\
	\mathbf{1}_{2} & 0
\end{array}\right), \quad \gamma^{0}=i \gamma^{5}=i\left(\begin{array}{cc}
	\mathbf{1}_{2} & 0 \\
	0 & -\mathbf{1}_{2}
\end{array}\right).
\end{eqnarray}
The $I_{a}\,(a=1,2,3)$ are the standard isospin matrices in some representation and we suppose that nothing depends on the extra coordinate $x^{4}$.

Let us now consider the supersymmetric quantum mechanical system defined by $(-1)^{F}=\gamma^{5}$, and
\begin{eqnarray}
H=\left(\frac{i}{\sqrt{2}} \slashed{D}_{4}\right)^{2}=\left(\begin{array}{cc}
	Q Q^{+} & 0 \\
	0 & Q^{+} Q
\end{array}\right)=\left(\begin{array}{cc}
	H_{1} & 0 \\
	0 & H_{0} \cdot \mathbf{1}_{2}
\end{array}\right).
\end{eqnarray}

Here
\begin{eqnarray}
H_{0}=\frac{1}{2}\left(\pi^{2}+\Phi^{2}\right), \quad H_{1}=H_{0} \cdot \mathbf{1}_{2}-\vec{\sigma} \cdot \vec{B}.
\end{eqnarray}
As a consequence of self-duality, the spin drops out of $Q^{+} Q$, while $Q Q^{+}$ contains the term $\vec{\sigma}\cdot\vec{B}$ characteristic of a particle
of ``apparent'' gyromagnetic ratio $4$. Note that $H_{0}$ is the Hamiltonian of a nonrelativistic, spinless particle moving in the field of the monopole.
 The Pauli Hamiltonian $H_{1}$ describes a nonrelativistic particle of spin $\frac{1}{2}$. For $E \neq 0$ the relation
\begin{eqnarray}
H_{1}=U^{+}\left(H_{0} \cdot \mathbf{1}_{2}\right) U, \quad U=\frac{1}{\sqrt{H_{0}}} Q^{+}
\end{eqnarray}
is satisfied. Thus $H_{1}$ and $H_{0} \cdot \mathbf{1}_{2}$ have identical spectra apart from zero modes which are counted
by the index theorem \cite{ref-5}. Let us now consider particles of definite momenta, helicity, and electric charge,
\begin{eqnarray}
q=\frac{I_{a} \Phi^{a}}{\|\Phi\|}\,,
\end{eqnarray}
scattering on the monopole. At large distances $A_{k}$ tends to the potential of a Dirac monopole, while the Higgs field behaves
as $\Phi \sim q\left(1-\frac{1}{r}\right)$. This implies that the transformation $U$ preserves the quantum numbers of the scattering states.
 Therefore $H_{1}$ and $H_{0} \cdot \mathbf{1}_{2}$ have identical $S$-matrices with a factorized, purely kinematical spin dependence.
 (The spin-charge factorization of the $S$-matrix is inherited  by the vector fluctuations scattering off the monopole \cite{ref-6}.) One can understand
 the factorization of the $S$-matrix also as a consequence of the conservation of $\vec{S}_{0}=\frac{1}{2}\left(\begin{array}{cc}0 & 0 \\ 0 & \vec{\sigma}\end{array}\right)$ for $H_{0} \cdot \mathbf{1}_{2}$ and
\begin{eqnarray}
\vec{S}_{1}=U^{+} \vec{S}_{0} U=\frac{-1}{2 H_{1}} \vec{\Omega}, \qquad \vec{\Omega}=\frac{1}{2}\left(\pi^{2}-\Phi^{2}\right) \vec{\sigma}+\Phi(\vec{\pi} \times \vec{\sigma})-(\vec{\sigma} \cdot \vec{\pi}) \vec{\pi}
\end{eqnarray}
for $H_{1}$. Now we introduce the following vector supercharges commuting with $H$ :
\begin{eqnarray}
Q_{1}=\frac{i}{\sqrt{2}} \slashed{D}_{4}=\left(\begin{array}{cc}
	0 & Q \\
	Q^{+} & 0
\end{array}\right), \quad Q_{2}=-i \gamma^{5} Q_{1}, \quad \vec{Q}_{\alpha}=2 i\left[\vec{S}_{0}, Q_{\alpha}\right], \quad(\alpha=1,2).
\end{eqnarray}
The even generators $\gamma^{5}, \vec{S}_{0}, \vec{S}_{1}$ and the odd ones $Q_{\alpha}, \vec{Q}_{\alpha}$ yield, for fixed $E \neq 0$, a closed superalgebra determined by the commutation relations:
\begin{subequations}
\begin{align}
	&
	{\left[S_{\alpha}^{i}, S_{\beta}^{j}\right]=i \delta_{\alpha \beta} \epsilon_{i j k} \quad(\alpha, \beta=0,1),}  \\
	&
	\left.\begin{array}{c}
	{\left[\gamma^{5}, Q_{\alpha}\right]=2 i \epsilon_{\alpha \beta} Q_{\beta}, \quad\left[\gamma^{5}, Q_{\alpha}^{k}\right]=2 i \epsilon_{\alpha \beta} Q_{\alpha}^{k} \quad(\alpha, \beta=1,2),} \\
	{\left[S_{0}^{k}, Q_{\alpha}\right]=-\frac{1}{2} i Q_{\alpha}^{k}, \quad\left[S_{0}^{k}, Q_{\alpha}^{j}\right]=\frac{1}{2} i\left(\delta_{j k} Q_{\alpha}+\epsilon_{k j n} Q_{\alpha}^{n}\right),} \\
	{\left[S_{1}^{k}, Q_{\alpha}\right]=\frac{1}{2} i Q_{\alpha}^{k}, \quad\left[S_{1}^{k}, Q_{\alpha}^{j}\right]=-\frac{1}{2} i\left(\delta_{j k} Q_{\alpha}-\epsilon_{k j n} Q_{\alpha}^{n}\right),}
    \end{array}\right\}  \\
    &
    \left.\begin{array}{c}
	\left\{Q_{\alpha}, Q_{\beta}\right\}=2 \delta_{\alpha \beta} H, \quad\left\{Q_{\alpha}, Q_{\alpha}^{k}\right\}=-4 H \epsilon_{\alpha \beta}\left(S_{0}^{k}+S_{1}^{k}\right), \\
	\left\{Q_{\alpha}^{k}, Q_{\beta}^{j}\right\}=2 H \delta_{\alpha \beta} \delta_{j k}-4 H \epsilon_{\alpha \beta} \epsilon_{k j n}\left(S_{1}^{n}-S_{0}^{n}\right).
    \end{array}\right\}
\end{align}\label{Eq.12}
\end{subequations}

For fixed $E>0$, this algebra is $S U(2 / 2) \oplus U(1)$. The $U(1)$ factor is generated by $\gamma^{5}$. Observe that $\vec{S}=\vec{S}_{0}+\vec{S}_{1}$ generates an $SO(3)$ invariance algebra of the Dirac operator $\slashed{D}_{4}$. The same algebraic structure is present (at least locally) in an instanton and in a $SD$ gravitational instanton background. This could be useful for analysing the Dirac equation.

Our investigations presented so far are valid in a general $SD$ background in the whole space. From now on we shall restrict ourselves to the behaviour at large distances. We show, in fact, that for the asymptotic, long range field of the $BPS$ 1-monopole there exists also a large dynamical symmetry \cite{ref-7}.  As $r \rightarrow \infty$, the fields tend to those of a singular monopole
\begin{eqnarray}
A_{i}^{a}=\epsilon_{\alpha i j} \frac{x_{j}}{r^{2}}, \quad \Phi^{a}=-\frac{x^{a}}{r}\left(1-\frac{1}{r}\right)\,. \label{Eq.13}
\end{eqnarray}
For the asymptotic system, the electric charge is conserved and the angular momentum (without spin) becomes
\begin{eqnarray}
\vec{L}_{0}=\vec{J}-\frac{\vec{\sigma}}{2}=\vec{r} \times \vec{\pi}-q \hat{r},
\end{eqnarray}
since $A_{i}^{a}$ describes an embedded Dirac monopole. The Hamiltonian $H_{0}$ is separated as
\begin{eqnarray}
H_{0}=-\frac{1}{2} \Delta_{r}+\frac{L_{0}^{2}-q^{2}}{2 r^{2}}+\frac{q^{2}}{2}\left(1-\frac{2}{r}+\frac{1}{r^{2}}\right)=-\frac{1}{2} \Delta_{r}+\frac{L_{0}^{2}}{2 r^{2}}-\frac{q^{2}}{r}+\frac{q^{2}}{2}\,.
\end{eqnarray}
The $\frac{q^{2}}{r^{2}}$ terms coming from the centrifugal potential and from $\Phi^{2}$ cancel and we are left with a Coulomb type system. This is the ``MIC-Zwanziger'' Hamiltonian \cite{ref-8,ref-9} introduced originally just for the sake of its symmetries $20$ years ago. It commutes with the Runge-Lenz vector
\begin{eqnarray}
\vec{K}_{0}=\frac{1}{2}\left(\vec{\pi} \times \vec{L}_{0}-\vec{L}_{0} \times \vec{\pi}\right)-q^{2} \hat{r}
\end{eqnarray}

Correspondingly, $\vec{L}_{1}=U^{+} \vec{L}_{0} U=\vec{J}-\vec{S}_{1}$ and $\vec{K}_{1}=U^{+} \vec{K}_{0} U$ are constants of motion for $H_{1}$. The Hamiltonian $H_{1}$ has been investigated recently by D'Hoker and Vinet \cite{ref-10} but without its present interpretation \cite{ref-11}. The spin $\frac{1}{2}$ ``Runge-Lenz vector'' $\vec{\Lambda}_{1}$ they have found is
\begin{eqnarray}
\vec{\Lambda}_{1}=\vec{K}_{1}-q \vec{S}_{1}=K_{0} \cdot \mathbf{1}_{2}+\vec{\pi} \times \vec{\sigma}+\left(\frac{q}{r}-\frac{q}{2}\right) \vec{\sigma}-\left(\vec{\sigma} \cdot \vec{B}\right)\vec{r}\,.
\end{eqnarray}
For bound states $0<E<\frac{q^{2}}{2}$, and from \eqref{Eq.12}, the angular momentum and the Runge-Lenz vector combine into an $[S U(2 / 2) \oplus U(1)] \oplus O(4)$ dynamical symmetry algebra of $H$. For $E=\frac{q^{2}}{2}$ and $E>\frac{q^{2}}{2}$ $O(4)$ is replaced by $E(3)$ and $O(3,1)$, respectively. The bound state spectrum (without zero modes)
\begin{eqnarray}
E=\frac{1}{2}\left( q^{2} -\frac{1}{n^{2}}\right), \quad n=|q|+1,|q|+2, \ldots
\label{Eq.18}
\end{eqnarray}
is calculated from the $O(4)$ symmetry by Pauli's method \cite{ref-12}. The multiplicity (for fix $q \neq 0)$ is $2\left(n^{2}-q^{2}\right)$. The $S$-matrix is also easy to derive from the bosonic symmetry $O(3) \oplus O(3,1)$ (the $O(3)$ factor comes from $\vec{S}$ ). In fact, Zwanziger's method \cite{ref-8} yields
\begin{eqnarray}
\begin{aligned}
	S\left(\underline{k}^{\prime}, s^{\prime} \mid \underline{k}, s\right) & =\delta\left(E_{k^{\prime}}-E_{k}\right) \cdot D_{s^{\prime}, s}^{\frac{1}{2}}\left(R_{\underline{k^{\prime}}}^{-1} R_{\underline{k}}\right) \cdot S_{0}\left(\underline{k}^{\prime} \mid \underline{k}\right),
\\[6pt]
	S_{0}\left(\underline{k}^{\prime} \mid \underline{k}\right) & =\sum_{l \geq|q|}(2 l+1)\left[\frac{\left(l-\frac{i}{k}\right) !}{\left(l+\frac{i}{k}\right) !}\right] D_{-q, q}^{l}\left(R_{\underline{k}^{\prime}}^{-1} R_{\underline{k}}\right)\,.
\end{aligned}
\end{eqnarray}
Here $\underline{k}$ is the velocity and $s$ is the helicity of the incoming particles of charge $q$, $k=\sqrt{2 E_{k}-q^{2}}$. The primed quantities refer to the outgoing particles. $R_{\underline{k}}$ is some rotation, chosen by convention, which brings the $\hat{z}$-direction into the direction of $\underline{k}$. The poles of $S$ are at $l-\frac{i}{k}=-T$, $T=1,2, \ldots$ yielding the $E>0$ bound state spectrum \eqref{Eq.18} with $n=T+l=|q|+1,|q|+2, \ldots$ .

Finally we show that, for the singular monopole \eqref{Eq.13}, $H_{0} \cdot \mathbf{1}_{2}$ and $H_{1}$ are essentially identical. In terms of the quantities
\begin{eqnarray}
x=\vec{\sigma} \cdot \vec{J}-\frac{1}{2}, \quad H_{r}=-\frac{1}{2} \Delta_{r}-\frac{q^{2}}{r}+\frac{q^{2}}{2}\,,
\end{eqnarray}
$H_{0} \cdot \mathbf{1}_{2}$ is rewritten as
\begin{eqnarray}
H_{0} \cdot \mathbf{1}_{2}=H_{r}+\frac{J^{2}+\frac{1}{4}-x}{2 r^{2}}\,.
\label{Eq.21}
\end{eqnarray}
The operator $x$ commutes with $H_{0} \cdot \mathbf{1}_{2}$ and with $\vec{J}$ and has eigenvalues $\pm\left(j+\frac{1}{2}\right)$ because of the identity
\begin{eqnarray}
x^{2}=J^{2}+\frac{1}{4}\,.
\end{eqnarray}
Thus, by separating according to $J^{2}$, $J_{3}$ and $x$ \eqref{Eq.21} becomes
\begin{eqnarray}
H_{0} \cdot 1_{2}=H_{r}+ \begin{cases}\frac{\left(j-\frac{1}{2}\right)\left(j+\frac{1}{2}\right)}{2 r^{2}} , & j=|q|+\frac{1}{2} , \ldots
\\[6pt]
\frac{\left(j+\frac{1}{2}\right)\left(j+\frac{3}{2}\right)}{2 r^{2}}, & j=|q|-\frac{1}{2}, \ldots\end{cases}
\quad .
\label{Eq.23}
\end{eqnarray}
On the other hand, the operator
\begin{eqnarray}
y=\vec{\sigma}\left(\vec{J}+2 q \hat{r}\right)-\frac{1}{2}
\end{eqnarray}
allows us to rewrite $H_{1}=H_{0} \cdot \mathbf{1}_{2}-\vec{\sigma} \cdot \vec{B}$ as
\begin{eqnarray}
H_{1}=H_{r}+\frac{J^{2}+\frac{1}{4}-y}{2 r^{2}}\,. \label{Eq.25}
\end{eqnarray}
Moreover, $y$ commutes with $H_{r}$ and $\vec{J}$ and has eigenvalues $\pm\left(j+\frac{1}{2}\right)$. Therefore, by diagonlizing the commuting set $J^{2}$, $J_{3}$, $y$ \eqref{Eq.25} is converted into
\begin{eqnarray}
H_{1}=H_{r}+ \begin{cases}\frac{\left(j-\frac{1}{2}\right)\left(j+\frac{1}{2}\right)}{2 r^{2}}, & j=|q|-\frac{1}{2}, |q|+\frac{1}{2}, \ldots
\\[6pt]
 \frac{\left(j+\frac{1}{2}\right)\left(j+\frac{3}{2}\right)}{2 r^{2}}, & j=|q|-\frac{1}{2}, \ldots\end{cases}\quad .
\end{eqnarray}
This is the same equation as \eqref{Eq.23} describing $H_{0} \cdot \mathbf{1}_{2}$! The associated eigenvalue problem can be solved in terms of hypergeometric functions. The result is consistent with \eqref{Eq.18}. The only difference between $H_{0} \cdot \mathbf{1}_{2}$ and $H_{1}$ is that for $H_{1}$, in the $y=\left(j+\frac{1}{2}\right)$ sector $j$ starts from $|q|-\frac{1}{2}$, while for $H_{0} \cdot \mathbf{1}_{2}$, in the $x=\left(j+\frac{1}{2}\right)$ sector $j$ starts from $|q|+\frac{1}{2}$. This is the only effect produced by the term $\vec{\sigma} \cdot \vec{B}$. In fact, this leads to the spectral asymmetry required by the index theorem \cite{ref-5}. The zero modes of $H_{1}$ carry the ``unpaired'' quantum numbers $j=|q|-\frac{1}{2}$, $y=|q|$. For fixed $q \neq 0$, the multiplicity of the $0$-energy state is $\left(2 j_{\min }+1\right)=2|q|$.

\bigskip\noindent\underline{Note added in 2024}. {\it This  paper was originally presented by L. Feher at the
{\sl  Symmetries in Science III} meeting at Schloss Hofen
 (July 25-28, 1988)} \cite{Feher:1988ym}.


\begin{thebibliography}{99}

\bibitem{ref-1} For a review, see
P.~Goddard and D.~I.~Olive,
``New Developments in the Theory of Magnetic Monopoles,''
Rept. Prog. Phys. \textbf{41} (1978), 1357
doi:10.1088/0034-4885/41/9/001

\bibitem{ref-2}
L.~S.~Brown, R.~D.~Carlitz and C.~k.~Lee,
``Massless Excitations in Instanton Fields,''
Phys. Rev. D \textbf{16} (1977), 417-422
doi:10.1103/PhysRevD.16.417

\bibitem{ref-3}
R.~Jackiw and C.~Rebbi,
``Spin from Isospin in a Gauge Theory,''
Phys. Rev. Lett. \textbf{36} (1976), 1116
doi:10.1103/PhysRevLett.36.1116

\bibitem{ref-4}
F.~A.~Bais and W.~Troost,
``Zero Modes and Bound States of the Supersymmetric Monopole,''
Nucl. Phys. B \textbf{178} (1981), 125-140
doi:10.1016/0550-3213(81)90499-5

\bibitem{ref-5}
E.~J.~Weinberg,
``Parameter Counting for Multi-Monopole Solutions,''
Phys. Rev. D \textbf{20} (1979), 936-944
doi:10.1103/PhysRevD.20.936

\bibitem{ref-6}
K.~J.~Biebl and J.~Wolf,
``Scattering of gauge and Higgs  fields by selfdual monopoles,''
Nucl. Phys. B \textbf{279} (1987), 571-592
doi:10.1016/0550-3213(87)90011-3

\bibitem{ref-7}
J. F. Sch\"onfeld,
``Dynamical Symmetry and Magnetic Charge,''
J. Math. Phys. \textbf{21} (1980), 2528
doi:10.1063/1.524360,
 L.~G.~Feher,
``Dynamical O(4) Symmetry in the Asymptotic Field of the Prasad-sommerfield Monopole,''
J. Phys. A \textbf{19} (1986), 1259-1270
doi:10.1088/0305-4470/19/7/026,
G.~W.~Gibbons and P.~J.~Ruback,
``The Hidden Symmetries of Multicenter Metrics,''
Commun. Math. Phys. \textbf{115} (1988), 267
doi:10.1007/BF01466773

\bibitem{ref-8}
D.~Zwanziger,
``Exactly soluble nonrelativistic model of particles with both electric and magnetic charges,''
Phys. Rev. \textbf{176} (1968), 1480-1488
doi:10.1103/PhysRev.176.1480

\bibitem{ref-9}
H.~V.~Mcintosh and A.~Cisneros,
``Degeneracy in the presence of a magnetic monopole,''
J. Math. Phys. \textbf{11} (1970), 896-916
doi:10.1063/1.1665227

\bibitem{ref-10}
E.~D'Hoker and L.~Vinet,
``Constants of motion for a spin 1/2 particle in the field of a dyon,''
Phys. Rev. Lett. \textbf{55} (1985), 1043-1046
doi:10.1103/PhysRevLett.55.1043,
E.~D'Hoker and L.~Vinet,
``Hidden Symmetries and Accidental Degeneracy for a Spin 1/2 Particle in the Field of a Dyon,''
Lett. Math. Phys. \textbf{12} (1986), 71
doi:10.1007/BF00400305

\bibitem{ref-11}
L.~G.~Feher and P.~A.~Horvathy,
``Non-relativistic scattering of a spin-1/2 particle off a self-dual monopole,''
Mod. Phys. Lett. A \textbf{3} (1988), 1451-1460
doi:10.1142/S0217732388001744
[arXiv:0903.0249 [hep-th]].

\bibitem{ref-12}
W.~Pauli,
``\"Uber das Wasserstoffspektrum vom Standpunkt der neuen Quantenmechanik,''
Z. Phys. \textbf{36} (1926) no.5, 336-363
doi:10.1007/BF01450175

\bibitem{Feher:1988ym}
L.~Feher, P.~Horvathy and L.~O'Raifeartaigh,
``Hidden symmetries of a self-dual monopole,''
In: {\sl  Symmetries in Science III,} Proc. Schloss Hofen
 Meeting, Gruber and Iachello (eds.), pp.525-529. Plenum (1989).


\end{thebibliography}
\end{document}